\documentclass[twocolumn]{jpsj2} 
\usepackage{graphicx,color}

\title{%
Degradation of superconductivity and spin fluctuations by electron over-doping in LaFeAsO$_{1-x}$F$_{x}$
}

\author{%
Shuichi \textsc{Wakimoto}$^{1,2}$,
Katsuaki \textsc{Kodama}$^{1,2}$,
Motoyuki \textsc{Ishikado}$^{1,2}$,
Masaaki \textsc{Matsuda}$^{1,2}$,
Ryoichi \textsc{Kajimoto}$^{2,3}$,
Masatoshi \textsc{Arai}$^{2,3}$,
Kazuhisa \textsc{Kakurai}$^{1,2}$,
Fumitaka \textsc{Esaka}$^{4}$,
Akira \textsc{Iyo}$^{2,5}$,
Hijiri \textsc{Kito}$^{2,5}$,
Hiroshi \textsc{Eisaki}$^{2,5}$,
Shin-ichi \textsc{Shamoto}$^{1,2}$,
}

\inst{%
$^{1}$ Quantum Beam Science Directorate, Japan Atomic Energy Agency,
   Tokai, Ibaraki 319-1195\\
$^{2}$ JST, Transformative Research-Project on Iron Pnictides (TRIP), 
   Tokyo 102-0075\\
$^{3}$ J-PARC Center, Japan Atomic Energy Agency,
   Tokai, Ibaraki 319-1195\\
$^{4}$ Nuclear Science and Engineering Directorate, Japan Atomic Energy Agency,
   Tokai, Ibaraki 319-1195\\
$^{5}$ Nanoelectronics Research Institute, National Institute of Advanced
   Industrial Science and Technology, Tsukuba, Ibaraki 305-8562
}

\recdate{\today}

\abst{%
Low energy spin fluctuations are studied for the electron-doped Fe-based superconductor LaFeAsO$_{1-x}$F$_{x}$ by inelastic neutron scattering up to the energy transfer of $\omega = 15$~meV using polycrystalline samples.  Superconducting samples ($x=0.057, T_c=25$~K and $x=0.082, T_c=29$~K) show dynamical spin susceptibility $\chi"(\omega)$ 
almost comparable with the parent sample's.  
However $\chi"(\omega)$ is almost vanished in the $x=0.158$ sample where the superconductivity is highly suppressed.
These results are compatible with the theoretical suggestions that the spin 
fluctuation plays an important role 
for the superconductivity.
}

\kword{%
Superconductivity, Spin fluctuation, Inelastic neutron scattering, Iron pnictide superconductor, LaFeAsO$_{1-x}$F$_{x}$.
}

\begin{document}
\maketitle

\section{Introduction}

Magnetic fluctuation has been expected to be a candidate as the origin of the Cooper pair formation in the high transition temperature (high-$T_c$) superconductivity since it appears with antiferromagnetic (AF) instability in the high-$T_c$ cuprates.~\cite{Birgeneau_JPSJ06}$^)$  Strikingly, recently-discovered new class of Fe-pnictide high-$T_c$ superconductor LaFeAsO$_{1-x}$F$_{x}$~\cite{Kamihara_08}$^)$ and family compounds show superconductivity just beside the AF regime in the $T-x$ phase diagram.~\cite{Huang_08,Chen_epl09}$^)$ 
%
Moreover the AF spin fluctuations have been observed by neutron scattering below $T_c$ for the superconducting 122~\cite{Christianson_08,Lumsden_09,Chi_09,Li_PRB09,Christianson_09}$^)$ and 11 compounds.~\cite{Mook_09,Qiu_09}$^)$
%
This similarity has drawn much attention to the new Fe-pnictide superconductors which give a unique opportunity to study the correlation between the AF spin fluctuation and the high-$T_c$ superconductivity.

LaFeAsO, a parent compound of the 1111-type Fe-pnictide superconductors, is an AF metal.  Band calculations indicate there are cylindrical Fermi surfaces of holes and electrons at $\Gamma$- and M-points, respectively.~\cite{Singh_08}$^)$  
Nesting between them induces 2-dimensional (2D) AF spin fluctuations that have been observed by inelastic neutron scattering near ${\rm\bf Q}^{2D}_{AF} = (1/2, 1/2, 0) = 1.10$~\AA$^{-1}$ in the tetragonal notation.~\cite{Ishikado_09}$^)$  Then, a 3-dimensional (3D) AF order develops below $T_N = 137$~K with AF propagation vector ${\rm\bf Q}^{3D}_{AF} = (1/2, 1/2, 1/2)$.~\cite{Cruz_08}$^)$
Substitution of oxygens by fluorine atoms and/or introducing oxygen vacancies provides electrons into the system.  The AF order is suppressed by $~\sim 4$~\% of F-doping and beyond this doping level the system shows superconductivity (Fig. 1(a)).

\begin{figure}[t]
\begin{center}
\includegraphics[width=8cm]{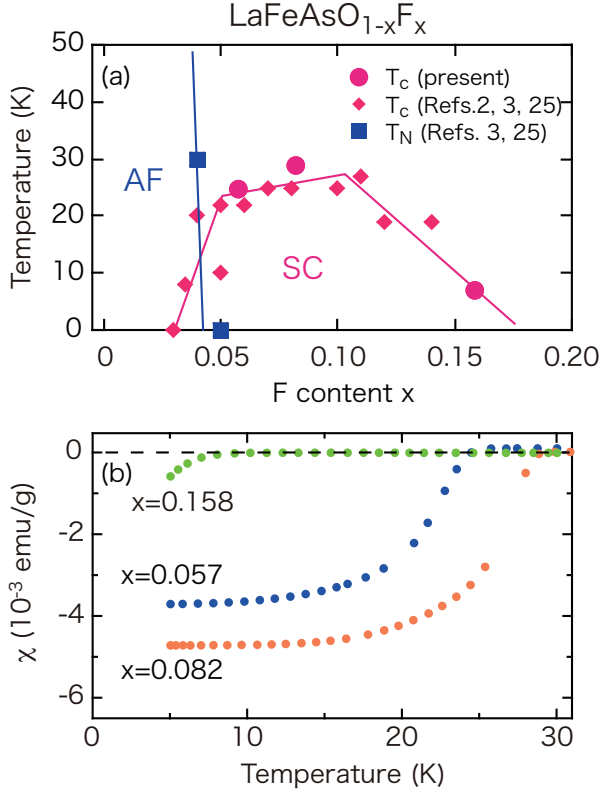}
\end{center}
\caption{(Color online) (a) $x-T$ phase diagram of LaFeAsO$_{1-x}$F$_{x}$.  Circles represent $T_c$ of the present samples.  Diamonds and squares indicate $T_c$ and $T_N$, respectively, adopted from previous works.~\cite{Kamihara_08,Huang_08,Nakai_JPSJ08}$^)$  (b) Meissner signals of the present polycrystalline samples measured in a cooling process under a magnetic field of 5~Oe.}
\end{figure}

From the early stage of the Fe-based superconductors research, many authors have pointed out the importance of spin fluctuations arising from the Fermi surface nesting in realizing the superconductivity.~\cite{Mazin_08,Cvetkovic_09,Kuroki_08,Cao_08,Ma_08,Yild_08}$^)$
Although the spin fluctuations have been observed in the superconducting 122 and 11 compounds, study of spin fluctuations of the 1111 system is very sparse due to the difficulty in synthesizing high quality samples.
The nuclear magnetic resonance (NMR) study of the LaFeAsO$_{1-x}$F$_{x}$ system shows that the spin fluctuations near $\omega=0$ dramatically decrease as doping increases up to $x=0.10$ whereas the $T_c$ changes only little.~\cite{Nakai_JPSJ08}$^)$  Apparently this behavior indicates weak coupling between the spin fluctuation and the superconductivity.
To reconcile these facts and clarify if the spin fluctuation plays a crucial role, a systematic study of spin fluctuations is desirable for the LaFeAsO$_{1-x}$F$_{x}$ system up to the overdoped region where the superconductivity is suppressed.  
%
For above purpose, we have performed inelastic neutron scattering on LaFeAsO$_{1-x}$F$_{x}$ with x=0.057 ($T_c=25$~K), $x=0.082$ ($T_c=29$~K), and overdoped $x=0.158$ (superconductivity is highly suppressed).

\section{Experimental details}

Powder samples of LaFeAsO$_{1-x}$F$_{x}$ have been synthesized by solid state reaction starting with nominal compositions of $x=0.05, 0.10$, and 0.20.  The $x$ values of the synthesized samples were determined by secondary ion-microprobe mass spectrometry to be 0.057(3), 0.082(5), and 0.158(7), respectively.
Powder x-ray diffraction data show that our samples contain only single 1111 phase with space group of $P4/nmm$ ($a=4.0$~\AA, $c=8.7$~\AA), demonstrating the high quality of the samples.  
Superconductivity of the prepared samples was characterized by SQUID measurements.  Figure 1(b) indicates Meissner signals for the three samples measured in a cooling process under a magnetic field of 5~Oe.  $T_c$ is characterized as an onset temperature of the Meissner signal and plotted in Fig. 1(a) by circles together with $T_c$ and $T_N$ reported in previous works.  $T_c$ of the $x=0.057$ and 0.082 samples agree well with previous reports.  By neutron diffraction we confirmed that all samples exhibit no AF order down to 4~K.  
The overdoped $x=0.158$ sample shows $T_c=7$~K, nevertheless it has a low volume fraction of about $10$~\% at 5 K.  Thus, the superconductivity of this sample is highly suppressed.

Inelastic neutron scattering experiments were performed using the triple-axis spectrometer TAS-1 installed at the research reactor JRR-3 of Japan Atomic Energy Agency.  Powder samples of $\sim 25$~g for each composition were 
used.  Collimation sequence of open-80$'$-S-80$'$-80$'$ (S denotes sample) and fixed final neutron energy at $E_f=30.5$~meV were utilized.  Inelastic measurements were done on the neutron energy loss condition.  This configuration gives instrumental resolutions of 3.5~meV in energy and 0.06~\AA$^{-1}$ in momentum transfer.
Volume ratios were estimated by nuclear Bragg intensities at $(0,0,2)$ normalized to $x$-dependent structure factor.~\cite{Nomura_08}$^)$  So obtained volume ratios of the samples with $x=0.057$, 0.082, and 0.158 were 1.0: 0.8 : 1.1.

\begin{figure}[t]
\begin{center}
\includegraphics[width=8cm]{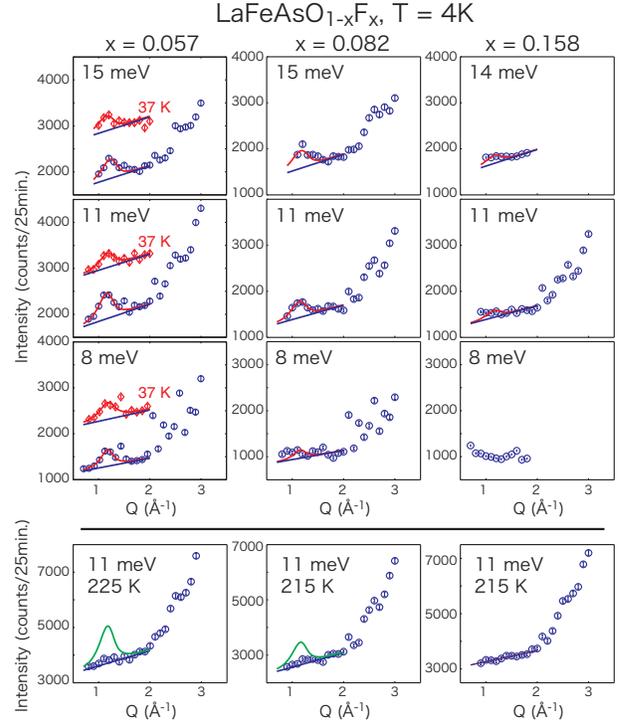}
\end{center}
\caption{(Color online) Neutron scattering profiles scanned as a function of momentum transfer $Q$ with fixed energy transfers $\omega = 8, 11$, and $15$~meV.  The top 9 panels show data at 4K, and the bottom 3 panels indicate data of $\omega = 11$~meV at $\sim 220$~K.  In the data panels of the $x=0.057$ sample, data at 37~K ($> T_c$) are also shown.  They are shifted by 1000 counts for clarity.  Solid lines in the top panels are fits to a resolution-convoluted Lorentzian function on sloped background.  Solid lines in the bottom panels show the expected intensity in case that the intensity at 4~K is phonon.}
\end{figure}

\section{Results}

Figure 2 shows neutron scattering intensity as a function of momentum transfer $Q$.  The intensity increases as $Q$ increases due to the $Q^2$-dependence of phonon scattering.  We focus on the expected magnetic scattering near $Q^{2D}_{AF}=1.1$~\AA$^{-1}$.
Data of $x=0.057$ at 37~K which is above $T_c$ are also shown.
Solid lines of the 4~K and 37~K data are the results of fits of the data in the range of $Q \leq 2$~\AA$^{-1}$ to a resolution-convoluted Lorentzian function.  Since the magnetic excitation is very steep against $Q$ in the energy range of $\omega \leq 15$~meV,~\cite{Ishikado_09}$^)$ we assumed the magnetic peak position $Q_{AF}$ to be independent of $\omega$.  The background level, which comes from mostly phonon contribution, is also adjusted as a sloped background. (Adjusted backgrounds are also shown in Fig. 2.)

It is shown that the $x=0.057$ sample shows clear peaks at all energies at both 4~K and 37~K.  
Data at 11~meV shows clear enhancement at 4~K.  
Existence of the magnetic peaks at 37~K evidences that the superconducting $x=0.057$ sample has spin fluctuations even above $T_c$.
The $x=0.082$ sample also shows peaks at 11 and 15~meV, although the peak structure at 8~meV is somewhat unclear.  The fittings for all these peaks give $Q_{AF} \sim 1.15$~\AA$^{-1}$ which is close to $Q^{2D}_{AF}$ and consistent with that observed for the parent compound for $T>T_N$.~\cite{Ishikado_09}$^)$  These peaks are gone at high temperatures.  Data at 11~meV at $T \sim 220$~K are shown as representative high-$T$ data in the lower panels of Fig. 2.  
The solid lines are peak profiles calculated by assuming that the peak observed at 4K is phonon contribution: that is, it depends on the temperature by $n(\omega)+1$, where $n(\omega)$ is the Bose factor $(e^{\omega/k_{B}T}-1)^{-1}$.  The profiles can not reproduce the observed data at all, indicating that
the observed peaks are indeed magnetic.

In contrast with these two samples, the overdoped $x=0.158$ sample shows no clear magnetic peaks at all energies at both low and high temperatures.  The fitting procedure to the $x=0.158$ data with float parameters results in a nearly zero magnetic intensity.  The solid lines of the 4K data are the results of fits by assuming the same background, peak position and width as those of $x=0.082$ to estimate maximum intensity of magnetic scattering.  Nevertheless, the intensity is still very small.
These facts evidence that the $x=0.158$ sample has no spin fluctuations in this energy range.

\begin{figure}[t]
\begin{center}
\includegraphics[width=8cm]{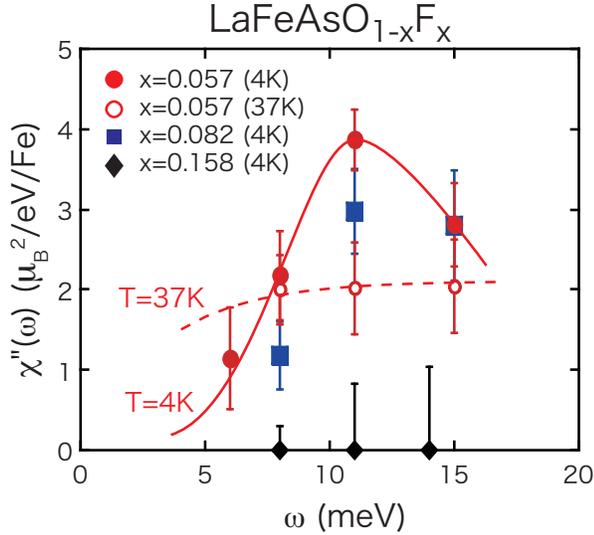}
\end{center}
\caption{(Color online) Imaginary part of dynamical spin susceptibility $\chi''(\omega)$ at 4~K calculated by normalizing to an incoherent scattering of standard vanadium.  Data at 37~K ($>T_c$) of the $x=0.057$ sample are also shown by open circles.  The error bars of data for only $x=0.158$ represent maximum values estimated by fitting with fixed backgrounds that are the same as those of $x=0.082$.  The solid and dashed lines are a guide to the eyes.}
\end{figure}

We have calculated absolute values of $Q$-integrated $\chi''(\omega)$ at 4~K by normalizing the magnetic cross sections to the incoherent scattering cross section of standard vanadium.  Results are summarized in Fig. 3.  
For the data of $x=0.158$, error bars represent the maximum values of $\chi''(\omega)$ estimated from the fore-mentioned fitting.
The superconducting samples of $x=0.057$ and 0.082 have maximum at $\omega \sim 11$~meV at 4~K, corresponding to $\sim 4.7 k_BT_c$.  Comparison to the spectrum of $x=0.057$ at 37~K clarifies that the maximum appears due to the enhancement below $T_c$.
For the $x=0.158$ sample, the $\chi''(\omega)$ is suppressed even at $11$~meV.  This demonstrates the suppression of the spin fluctuations in the energy range up to 15~meV.
We summarize $x$-dependence of $\chi''(\omega)$ in Fig. 4.  In the figures of $\omega=8$ and 11~meV, $\chi''(\omega)$ of the parent compound are also presented by open symbols.  These are measured with the same spectrometer configuration using the identical sample reported in Ref.~\citen{Ishikado_09}).
The spin fluctuations in the $x=0.158$ sample is highly suppressed, whereas those of the superconducting $x=0.057$ and $0.082$ samples are comparable to those of the parent compound.  Thus, an appreciable amount of spin fluctuations survives in the energy range of $\omega \leq 15$~meV in the superconducting samples.

\begin{figure}[t]
\begin{center}
\includegraphics[width=8cm]{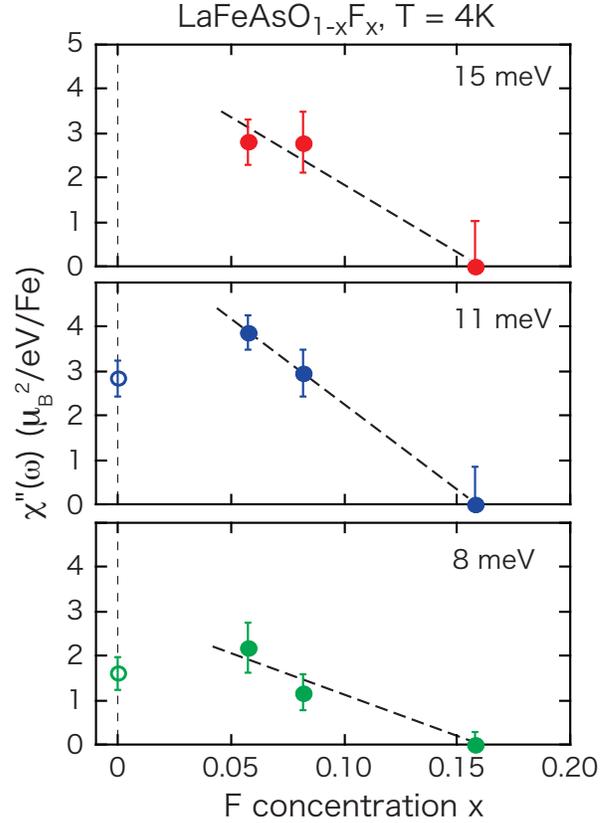}
\end{center}
\caption{(Color online) Imaginary part of dynamical spin susceptibility $\chi''(\omega)$ at $\omega = 8, 11$, and 15~meV as a function of F content $x$.  Open symbols show $\chi''(\omega)$ of the parent compound measured at just above the N\'{e}el temperature 140~K, where the $\chi''(\omega)$ is maximum.}
\end{figure}

\section{Discussion}

We have shown that the spin fluctuation in LaFeAsO$_{1-x}$F$_{x}$ becomes suppressed by electron-overdoping at 4~K.  
%
It is reported that the superconducting Fe-122 and 11 systems show the enhancement of the spin fluctuation at energy transfer of $4.2 \sim 5.3$~$k_BT_c$ below $T_c$ observed by neutron scattering.~\cite{Christianson_08,Lumsden_09,Chi_09,Li_PRB09,Christianson_09,Mook_09,Qiu_09}$^)$
%
We have observed qualitatively similar enhancement for the present 1111 sample with $x=0.057$ at 11~meV ($\sim 5.1$~$k_BT_c$).  
This feature has been explained by either $s_{\pm}$ scenario due to the superconducting gap symmetry,~\cite{Kors_08,Maier_09}$^)$ or simple $s_{++}$ scenario due to the redistribution of spectral weight by the gap opening.~\cite{Onari_09}$^)$  Distinguishing these two requires more detailed measurements and which is not the main scope of this Letter.~\cite{shamoto_10}$^)$  
Instead we put emphasis on the suppression of the magnetic fluctuation in the overdoped sample.

The present $x=0.057$ sample shows well defined spin fluctuation above $T_c$ demonstrating the existence of the bare spin fluctuations without the enhancement below $T_c$.  The disappearance of the magnetic signal in $x=0.158$ at 4~K evidences the disappearance of the bare spin fluctuations.
It is reasonable that the hole Fermi surface at the $\Gamma$-point shrinks by electron doping and eventually disappears by over-doping.  This results in a poor nesting condition to the electron Fermi surface at the M-point and a suppression of the spin fluctuations.  
Recent ARPES measurements on BaFe$_{2-x}$Co$_{x}$As$_{2}$, where the Co-doping supplies electrons, suggest a bad nesting condition due to the shrinkage of the hole Fermi surface in a non-superconducting over-doped sample.~\cite{SatoARP_09}$^)$
%
In addition, neutron measurements of overdoped BaFe$_{2-x}$Co$_{x}$As$_{2}$ shows suppression of the inelastic magnetic scattering, consistent with our results.~\cite{Matan_09CM}$^)$

We cannot rule out the possibility that the magnetic scattering in the $x=0.158$ sample still exists at lower $Q$, which is not accessible in the present configuration.  However the scattering near $Q=1.1$~\AA$^{-1}$~ disappears clearly, which corresponds to the nesting vector from the $\Gamma$ to $M$ points.
These facts imply the importance of the Fermi surface nesting between the $\Gamma$ and $M$ points for the superconductivity in the Fe pnictides.  
The nesting should induce elementary fluctuations which may act as a source of the superconductivity, such as spin, orbital, and charge fluctuations.  So far, only the spin fluctuation has been observed to disappear at the overdoped regime.  Although this is not a direct evidence of spin driven superconductivity, but it is an implication of the coupling between the spin fluctuation and the superconductivity.

Finally, we mention a difference of the spin fluctuations probed by the present study and NMR.  Our neutron measurement shows that an appreciable amount of $\chi"(\omega)$ which is comparable to that of the non-doped sample is present in the superconducting samples with $x=0.057$ and $x=0.082$.  
In contrast, NMR-$1/T_{1}T$ measurement on this system revealed that the spin fluctuation in the low energy region which is much lower than that of our neutron measurement is suppressed by small doping.~\cite{Nakai_JPSJ08}$^)$
In our measurements, the peak structure clearly observed in $x=0.057$ at 8 meV becomes somewhat unclear in $x=0.082$.  Even lower energy region may show a drastic decrease of $\chi''(\omega)$ with doping.
Clearly the lower energy measurements of neutron scattering at temperature range above $T_c$ are necessary using single crystals.  
%

\section{Summary}

Systematic neutron scattering study of LaFeAsO$_{1-x}$F$_{x}$ revealed that the spin fluctuations up to 15~meV that are comparable to the non-doped LaFeAsO survive in the superconducting samples with $x=0.057$ and $0.082$, whereas they are highly suppressed in the over-doped $x=0.158$ where the superconductivity is also highly suppressed.  This can be understood by a disturbed nesting condition due to the reduction of the hole-Fermi surface at the $\Gamma$-point upon electron over-doping. 
%
Our observation is compatible with theoretical suggestion that the spin fluctuations due to the Fermi surface nesting is important for the iron-pnictide superconductivity.

\begin{acknowledgements}

Authors thank  M. Machida, A. Q. R. Baron, and T. Fukuda
for invaluable discussions.
This work is partially supported by a Grant-in-Aid for Specially Promoted Research 17001001 from the Ministry of Education, Culture, Sports, Science and Technology, Japan.

\end{acknowledgements}


\end{document}